\RequirePackage[2020-02-02]{latexrelease}

\documentclass[tighten]{emulateapj}

\slugcomment{Accepted, \today}

\shorttitle{Global Extinction}
\shortauthors{Steinbring}

\input epsf
\def\plotone#1{\centering \leavevmode
\epsfxsize=1.00\columnwidth \epsfbox{#1}}

\def\plotoneextratwiddle#1{\centering \leavevmode
\epsfxsize=0.97\columnwidth \epsfbox{#1}}
\def\plotoneextratwiddleagain#1{\centering \leavevmode
\epsfxsize=0.96\columnwidth \epsfbox{#1}}
\def\plottwocolumnwide#1{\centering \leavevmode
\epsfxsize=2.00\columnwidth \epsfbox{#1}}
\def\plottwocolumnwidetwiddle#1{\centering \leavevmode
\epsfxsize=1.99\columnwidth \epsfbox{#1}}
\def\plottwocolumnwidetwiddleagain#1{\centering \leavevmode
\epsfxsize=1.98\columnwidth \epsfbox{#1}}
\def\plottwocolumnmedium#1{\centering \leavevmode
\epsfxsize=1.65\columnwidth \epsfbox{#1}}
\def\plottwocolumnmediumtwiddle#1{\centering \leavevmode
\epsfxsize=1.64\columnwidth \epsfbox{#1}}

\def\plotonemedium#1{\centering \leavevmode
\epsfxsize=0.95\columnwidth \epsfbox{#1}}

\def\plottwocolumnfiddle#1#2{\centering \leavevmode
\epsfxsize=0.67\columnwidth \epsfbox{#1} \hfil \hspace{1.5 mm}
\epsfxsize=0.70\columnwidth \epsfbox{#2}}

\begin{document}

\title{Global Extinction: Gemini North and South GMOS Combined Photometry\\ Relative to the {\it Gaia} Catalog, and Long-Term Atmospheric Change}

\author{Eric Steinbring\altaffilmark{1}}

\altaffiltext{1}{National Research Council Canada, Herzberg Astronomy and Astrophysics, Victoria, BC V9E 2E7, Canada}

\begin{abstract} 
Effects of long-term atmospheric change were looked for in photometry employing Gemini North and South twin Multi-Object Spectrograph (GMOS-N and GMOS-S) archival data. The whole GMOS imaging database, beginning from 2003, was compared against the all-sky {\it Gaia} object catalog, yielding $\sim 10^6$ Sloan ${\rm r}^\prime$-filter samples, ending in 2021. These were combined with reported sky and meteorological conditions, and versus a simple model of the atmosphere plus cloud together with simulated throughputs. One exceptionally extincted episode in 2009 is seen, as is a trend (similar at both sites) of about $2~{\rm mmag}$ worsening attenuation per decade. This is consistent with solar-radiance transmissivity records going back over six decades, aerosol density measurements, and more than $0.2~^\circ{\rm C}/{\rm decade}$ rise in air temperature, which has implications for calibration of historic datasets or future surveys. 
\end{abstract}


\section{Introduction}\label{introduction}

\begin{figure*}
\plottwocolumnmedium{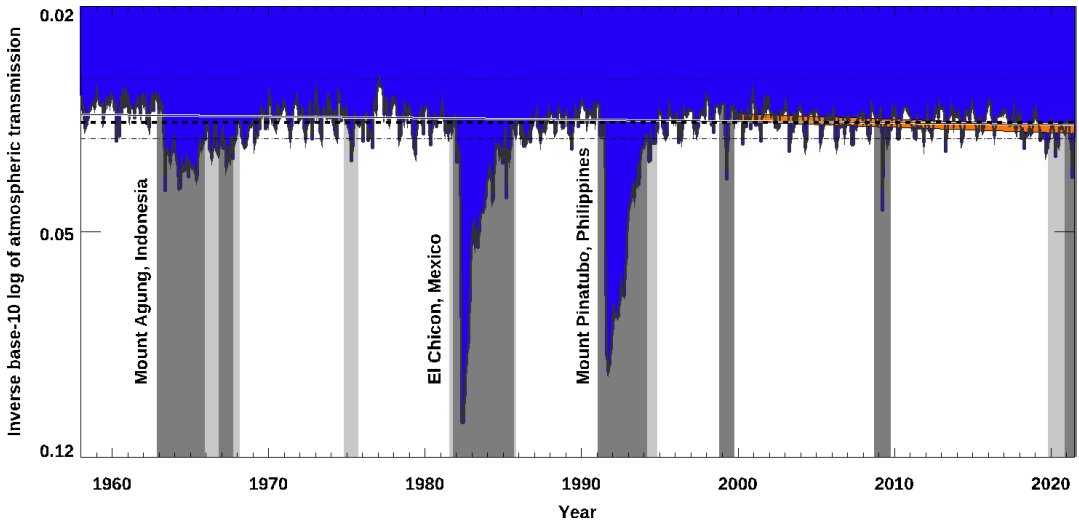}
\caption{Mean monthly atmospheric transmission via direct solar radiation; median is indicated by dashed line; dotted is maximum monthly-averaged recorded value; a thin black line shows a least-squares linear fit to all records (``greyed-out" times excluded); orange line is the same, restricted to just those since the year 2000.}
\label{figure_atmospheric_transmission_with_time}
\vspace{2.0 mm}
\end{figure*}

It is now broadly recognized that a warming global climate has impacts for astronomy \citep{Cantalloube2020, Flagey2021} which might include worsening seeing \citep{Sarazin2008} or increased, more-variable precipitable water vapour with rising air temperatures \citep{Bohm2020}. One outcome may be more nighttime dome closures due to bad weather \citep{vanKooten2022}. And at sites worldwide, poorer conditions - with their attendant harm to the quality of observations - are predicted from climate-change modeling \citep{Haslebacher2022}. Along with greater humidity, higher attenuation would be particularly detremental to future wide-field optical surveys employing precise photometry of faint, possibly transient variation: detectability depends directly on atmospheric transparency, setting both survey depth and the accuracy of zeropoints. 

While the zeropoint precision of a ground-based optical/infrared (OIR) telescope can be maintained by tracking the flux of external calibrators, its best-possible sensitivity (least difference from the ideal star-catalog value) declines either with the decay of transmission for any optic or when atmospheric transparency decreases. At issue for a large dataset -- spanning many years and pursuing percent-level photometry -- will be whether that decrease is detectable within those data next to changes in the optics and instrument itself. For a sense of the relative strength and timescale of effects: Abrupt efficiency gains come from detector upgrades, with improvements of perhaps $20\%$ or more. And even though detectors maintain nearly uniform sensitivity during their operational life of a decade or so, reflectivity of optical coatings typically degrade by up to about $2\%$ per year per mirror despite cleaning on intervals of several months to slow decay, which is regained when recoated. Against these changes, however, zeropoints might be measured weekly or less often via photometry of standard stars, in non-ideal seeing, under skies thought to be clear or cloud-free; then corrected to zenith airmass, and subtracted to achieve an instrumental throughput. Meanwhile, the transparency of the atmosphere fluctuates with pressure, temperature and humidity night-to-night at the $5\%$ level on top of seasonal variation. It also incurs other fluctuations in aerosols, including due to volcanic episodes, possibly occuring about once a decade and dropping $10\%$ in the optical, lasting months and seen worldwide (e.g. Miles, 1983; Dutton et al. 2011 and references therein): datasets spanning such periods would require particular care in their calibration, as sensitivity during that time will be correspondingly less. But a smaller, persistently growing decrement could erode sensitivity as well. Could a net drop in transparency of 1\% be detectable?

One useful dataset in which to look for such a loss is from the 8-meter Gemini telescopes, which saw first light in 1999 on Maunakea, Hawai'i (Gemini North: 19.8238 deg N, 155.469 deg W, elev. 4213 m a.s.l.), and in 2000 on Cerro Pachon, Chile (Gemini South: 30.2407 deg S, 70.737 deg W, elev. 2722 m a.s.l.). Their twin Multi-Object Spectrographs (GMOS-N, North; GMOS-S, South: Murowinski et al. 1998) were commissioned in 2003 \citep{Hook2004} and have remained the most-used instruments on each telescope. They include an imaging mode covering a $5.5~{\rm arcmin}\times 5.5~{\rm arcmin}$ square field of view, with Sloan Digital Sky Survey (SDSS) filters: ${\rm g}^\prime$, ${\rm r}^\prime$, ${\rm i}^\prime$ and ${\rm z}^\prime$ (${\rm u}^\prime$ is used rarely). The ${\rm r^\prime}$ filter is a good match to the G central wavelength onboard {\it Gaia}. Gemini South is also co-located near the site of the new Rubin Observatory, which will undertake the Legacy Survey of Space and Time (LSST) and be calibrated with separate, dedicated instrumentation \citep{Coughlin2018}. Encompassing the whole sky using two identical instruments allows uniform photometric analysis, and can serve together with an all-sky object catalog as a proxy to a single standard calibration field of sources for both.

In this study, 17 years of GMOS imaging data, meterological records, nightlogs, and reduction-pipeline zeropoints were used to track the year-to-year optical atmospheric attenuation for Gemini Observatory. First, in Section~\ref{modeling}, over 63 years of daytime solar-radiance transmission estimates are introduced, providing an independent reference for expected atmospheric changes. Together with aerosol absorption and weather-station data, a simple semi-empirical model of extinction by a warming atmosphere is developed, matching those. Thin, unseen cloud is incorporated by calculating a corrective offset. This approach has the merit of allowing the return of an atmospheric attenuation estimate based on the median difference of object magnitudes from their catalog values, which is achievable in essentially every GMOS frame, and so relative to the overall median extinction (and zeropoint, by definition) can be statistically robust. A numerical instrument simulation then guides a method to achieve the necessary data sampling and sensitivity to detect long-term changes in distribution. The archival images and corresponding {\it Gaia} catalog are presented in Section~\ref{analysis}, and photometric analyses are described. A worsening trend in attenuation at $0.6~\mu{\rm m}$ wavelength (within Sloan ${\rm r}^\prime$ band) of about $2~{\rm mmag}/{\rm decade}$ is found, consistent with a proportional increase in atmospheric temperature more than $0.2~^\circ{\rm C}/{\rm decade}$, constituting an effective $1\%$ loss in (combined) aperture since first light. One episode in 2009 approached $5\%$ beyond the median extinction of $0.11~{\rm mag}$. Finally, Section~\ref{summary} summarizes results and provides some considerations for synoptic surveys, and the potential impact due to climate change.

\section{Tracking Attenuation Differences}\label{attenuation}

Atmospheric transmission is routinely recorded during the day, when skies are clear, using solar radiation measurements at Mauna Loa Observatory (MLO), Hawai'i. These are obtained in visible light, most sensitive at $0.6~\mu{\rm m}$. They are already well-studied, looking for episodic (and punctuated) events related to solar and seasonal cycles; and after accounting for these, there was found to be a slow trend, consistent with up to $-0.15\%$ per decade change in transparency prior to 2000 \citep{Dutton1985, Dutton2001}, and possibly steepening afterwords \citep{Dutton2011}. 

All MLO transmission measurements through 2021 were downloaded from from the public NOAO website archive\footnote{https://gml.noaa.gov/webdata}, providing a uniform record that stretches back to 1958. The monthly averages are shown in Figure~\ref{figure_atmospheric_transmission_with_time}, with periods of low transmission shaded: light grey when those dipped more than one-third standard deviation below the overall mean value for at least three continuous months; dark-gray when that was by a full standard deviation for the same duration or more. Three such periods are notable, associated with known major volcanic eruptions. Even more severe impacts have occured before: that of Krakatoa in 1883 was still observed as a ``Bishop's Ring" around the Sun years later \citep{Backhouse1893}. At least one much-lesser instance has occurred since the start of Gemini operations, plausibly associated with the 22 March 2009 eruption of Mount Redoubt, in Alaska.  Also notably, since the end of 2019, monthly transmission had again fallen below the mean of these data, although that may not persist and it is not yet apparent what caused that. Even so, the previously reported worsening trend is confirmed: a least-squares fit to data after the ``greyed-out" periods are removed has slope of $-0.04\%$ per decade; a similar fit restricted to data post-2000 is indeed steeper at $-0.17\%$ per decade.

\begin{figure}
\hspace{2.5 mm}\plotoneextratwiddleagain{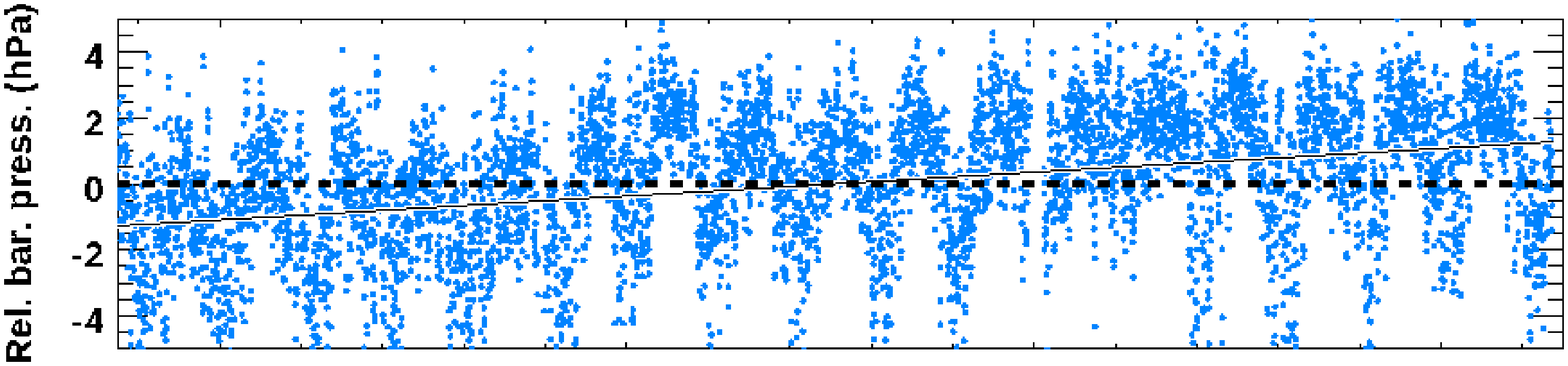}\\
\vspace{0.5 mm}
\hspace{2.5 mm}\plotoneextratwiddleagain{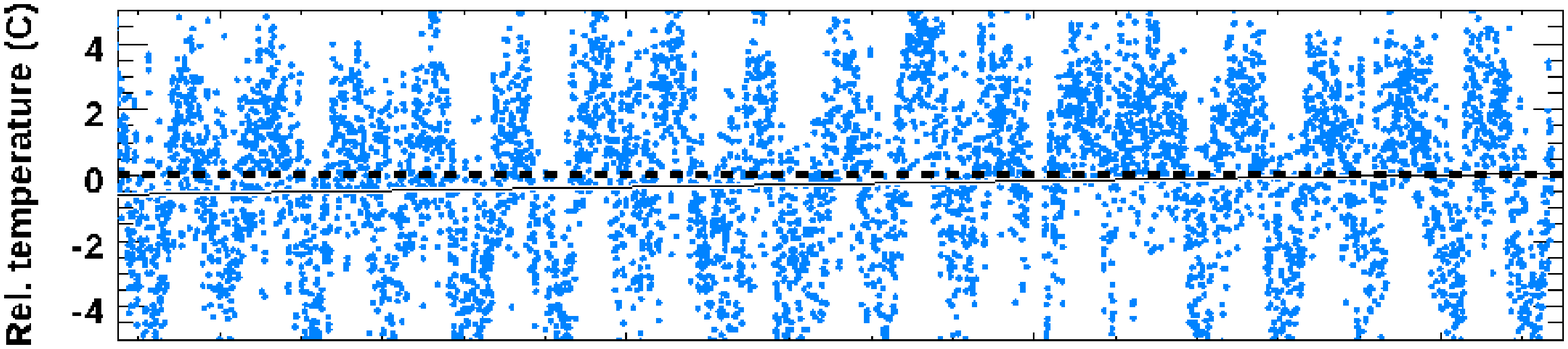}\\
\vspace{1.0 mm}
\hspace{1.25 mm}\plotoneextratwiddle{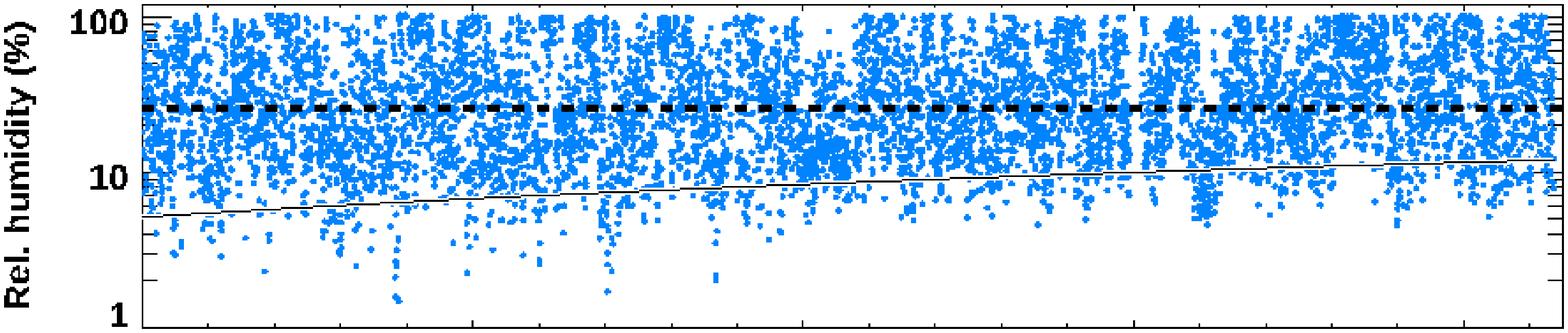}\\
\vspace{1.5 mm}
\plotone{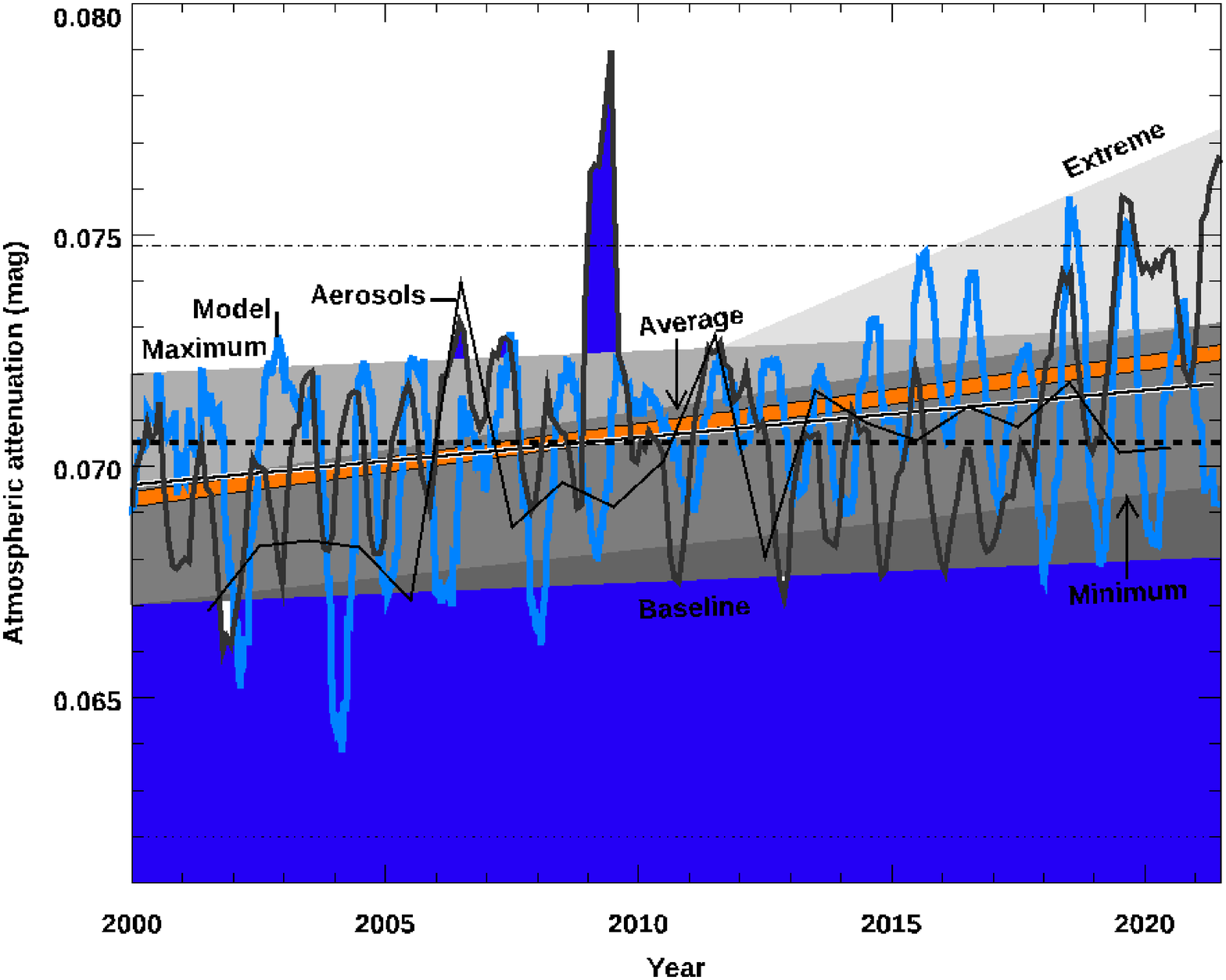}
\caption{Same as Figure~\ref{figure_atmospheric_transmission_with_time} post-2000 as atmospheric attenuation in magnitudes, with addition of nephelometer yearly averages (thin black curve). Contemporaneous CFHT weather-station data (top panels, blue dots) for barometric pressure, air temperature and relative humidity have each had their mean over this period subtracted. These data are inputs to the model as in Section~\ref{modeling} (blue curve), with the simpler limiting cases (minimum, average and maximum: increasing grey-shading; extreme:lightest-grey). Rising humidity with increased temperature matches the trend in attenuation; note ``worst" but unrelated period in 2009. }
\label{figure_attenuation_from_mlo_against_cfht_weather_data}
\end{figure}

Data after Gemini first light are replotted in Figure~\ref{figure_attenuation_from_mlo_against_cfht_weather_data} as the decrement in transmission, converted to magnitudes. A further comparison is relative to MLO nephelometer measurements of aerosol absorption, also available from the NOAA public archive. The yearly average values (in the $0.5~\mu{\rm m}$ channel) after 2000 are here normalized to the same median (decrement) of transmission to better compare them. These appear to show a similar trend, with that for absorptions at least as steep as obtained from direct transmission measurements. In the analysis that follows, it will be shown that this can be taken as tracking the change in atmospheric attenuation - when the effect of cloud is dealt with - as if those were under photometric skies, without any detectable cloud.

\subsection{Modeling Atmospheric Attenuation Change}\label{modeling}

A practical model of atmospheric attenuation is to treat air as a bulk absorber within the telescope beam, that is, it is linear with airmass $Z$, and so scales with surface pressure $p$ and relative humidity $\rho$ as $Z\propto p\rho/(\bar{p}\bar{\rho})$ \citep{Steinbring2009}. If those variables were truly Gaussian, with a symmetric variance, the mean (and median) would also define the modal or peak value as per $$\bar{Z}={1\over{\sqrt{2\pi}\sigma}}\int_0^\infty{\exp{{\Big[}-{\textstyle (Z-1)^2\over{2\sigma^2}}{\Big]}}}\,{\rm d}Z = 1, \eqno(1)$$ where $\sigma$ is the distribution standard deviation. The total extinction along the path should then be described by $$A = A_{\rm air} + A_{\rm cloud} = A_0 + (\Delta A)\times Z t, \eqno(2)$$ in magnitudes while clouds absent, that is, air only when $A_{\rm cloud}$ is zero and $\Delta A$ is the rate of change in atmospheric attenuation over time $t$ for $A_0$ as initial value. To clarify: the slope of change in atmospheric attenuation over long timescales can be found by assuming $A_0$ includes cloud. And to follow, in Section 2.2, the effect of that assumption will be dealt with by finding an offset from the median of $A$ which is due to very thin, persistent cloud.

To confirm the fidelity of this approach, nightly weatherstation data dating back to 1979 were obtained from the Canada-France-Hawaii Telescope (CFHT), which is nearby Gemini North - effectively the same elevation on the summit of Maunakea. These are continuous data, taken in all weather, not affected by dome-open conditions. These are shown in Figure~\ref{figure_attenuation_from_mlo_against_cfht_weather_data}: the blue curve is the model of equation 2 (normalized by the medians). Even more simply, replacing pressure with a linear relationship to surface air temperature $T$ via $0.661~{\rm hPA}~^\circ{\rm C}^{-1}$ (their average through 2021), and similarly, taking the associated rate of change in relative humidity at the mean temperature ($7.03\%$ at $-2.53~^\circ{\rm C}$ during this period) it can be reduced to a single variable: $$\Delta A = 2.39 \times \Delta T~{\rm mmag}/{\rm decade}. \eqno(3)$$ That is shown in Figure~\ref{figure_attenuation_from_mlo_against_cfht_weather_data} for $\Delta T=0.7~^\circ{\rm C}/{\rm decade}$ or, equivalently $\Delta A=1.7~{\rm mmag}/{\rm decade}$, corresponding to the average change at CFHT. Also shown is an encompassed range, chosen to reflect the uncertainty associated with fitting a starting point and slope due to fluctuations in attenuation per year and working backwards to obtain change in temperature. These prescribe ``maximum" and ``minimum" cases allowable including the standard deviation of $0.005~{\rm mag}$, plus an ``extreme" case of both (low) starting point and (high) slope limits. Table~\ref{table_models} lists these variants, labeled by increasing slope, along with $0.1~^\circ{\rm C}/{\rm decade}$, the post-industrial (1880 onwards) ``baseline" used in the Annual 2020 Global Climate Report \citep{NOAA2021} where $0.18~^\circ{\rm C}/{\rm decade}$ is the generally accepted 40-year average, since 1981. A caution is that this is for surface temperature at sea-level, not the mountain summits nor the upper atmosphere above them. But it is the simple scaling property of the model, integrated along the air column, that makes comparison with this rate a useful reference.

\begin{deluxetable}{lcccc}
\tablecaption{Atmospheric Model\label{table_models}}
\tablewidth{0pt}
\tabletypesize{\tiny}
\tablehead{\colhead{} &\colhead{} &\colhead{$A_0$} &\colhead{$\Delta T$} &\colhead{$\Delta A$}\\
\colhead{Label} &\colhead{Time Period} &\colhead{(mag)} &\colhead{($^\circ{\rm C}\over{\rm decade}$)} &\colhead{(${{\rm mmag}\over{\rm decade}}$)}}
\startdata
Baseline             &1958.0-2021.5  &0.0625 &0.1 &0.2\\
Minimum              &2000.0-2021.5  &0.0670 &0.2 &0.5\\
Average              &2000.0-2021.5  &0.0695 &0.7 &1.7\\
Maximum              &2000.0-2021.5  &0.0720 &1.3 &3.1\\
Extreme              &2000.0-2021.5  &0.0670 &2.0 &4.8
\enddata
\end{deluxetable}

\subsection{Accounting for a Shift in Distribution, or Cloud}\label{clouds}

\begin{figure}
\plotone{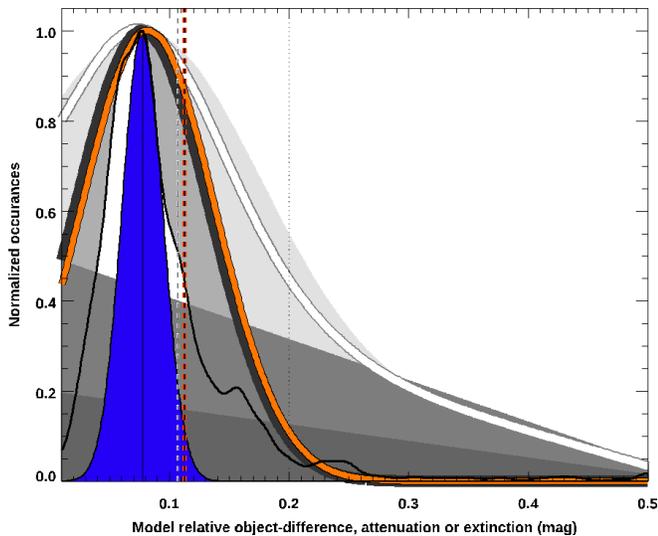}
\caption{Model distribution of attenuation (thick black curve) and that shifted $1\%$ (orange) for a Gaussian atmosphere and power-law clouds: total outer-envelope case (light grey shading) or $80\%$ observed photometric skies (dark grey); initial median extinction (vertical black, dashed line); and total (dashed red), offset from the distribution mode (thin black line) towards photometric limit (dotted). The resulting object magnitudes after subtracting their catalog value, as per Section~\ref{instrument} (thick white curve) and limiting ``residual" difference between {\it Gaia} and SDSS (outlined white region) relative to a $2\%$ photometric uncertainty (blue region).}
\label{figure_model_atmospheric_extinction_distribution}
\vspace{2.0 mm}
\end{figure}

When present, clouds are well described as a grey absorber with duration $\delta$ declining in occurence $N$ approximately with magnitude as $N\propto 1 - \alpha \log (\delta/\bar{\delta})$: thinnest cloud occurs most often, thick cloud is rare \citep{Steinbring2012}. A value of $\alpha=-1.84$ was found via CFHT SkyProbe data, for $\bar{\delta}=2.14~{\rm hr}$, giving a cloud-free fraction of $56\%$ with the dome open $75\%$ of the time \citep{Steinbring2009}. The effect is to reshape the distribution of attenuation, $A$, which is readily calculated via numerical integration without assuming {\it a priori} some value for $A_0$. This is illustrated in Figure~\ref{figure_model_atmospheric_extinction_distribution}. It has a median of ${\tilde A}=0.1070~{\rm mag}$ (medians hereafter designated by a tilde above symbol) and a peak of ${\hat A}=0.0763~{\rm mag}$ (the mode) using $A_{\rm air} = 0.0720~{\rm mag}$; a value of $\sigma=0.05$ is employed throughout, found to well bound the observed variance. Thus, increasing cloudiness offsets the median extinction in the same sense as growing atmospheric attenuation (here considering a 1\% shift) but less so if thin; a resulting total distribution with a photometric error of 2\% (added in quadrature) is indicated by the lightly-shaded ``envelope" region; a limiting case of cloud-free observed fraction of $80\%$ is also shown. When $0.2~{\rm mag}$ (dotted line) or less, cloud is undetectable visually, which are conditions generally considered ``photometric." 

The relative pressure altitude between Gemini North and MLO of $617/680\approx 0.91$ (means, in hPa) and that between MLO and Gemini South of $680/730\approx 0.93$ happen to be nearly equal. So a simplification is to choose a single (median) case to account for thin, unseen cloud under photometric skies at both sites. Choosing the maximum attenuation by air, as above, this amounts to a constant, maximal systematic median to modal offset of $$A_{\rm offset}={\tilde A}-{\hat A}\approx 0.0307~{\rm mag}. \eqno(4)$$ That is also conservative because the slightly-higher relative difference in offset towards higher pressure is still small ($\leq 0.005~{\rm mag}$) but can serve only to underestimate attenuation for Gemini South for similar cloudiness fractions. That remains true for more observations obtained from South than North, seen later to be the usual case.

\subsection{Simulating Observational Effects, and Sampling}\label{instrument}

\begin{figure*}
\hspace{0.5 mm}\plottwocolumnmediumtwiddle{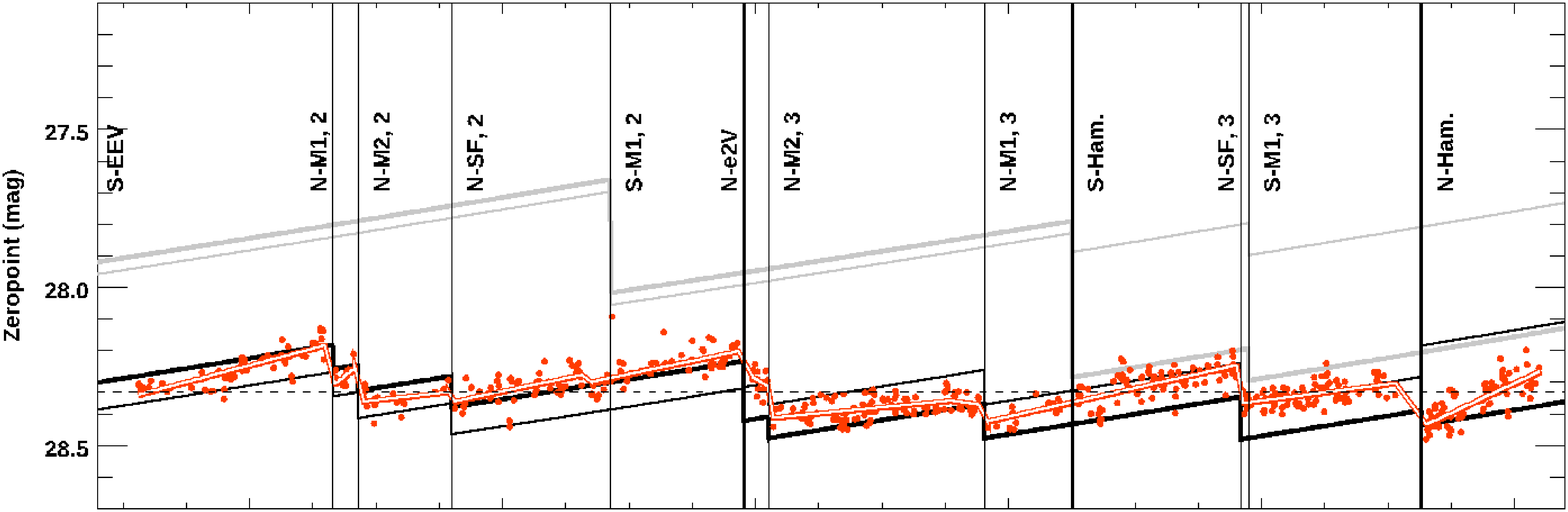}\\
\vspace{1.0 mm}
\plottwocolumnmedium{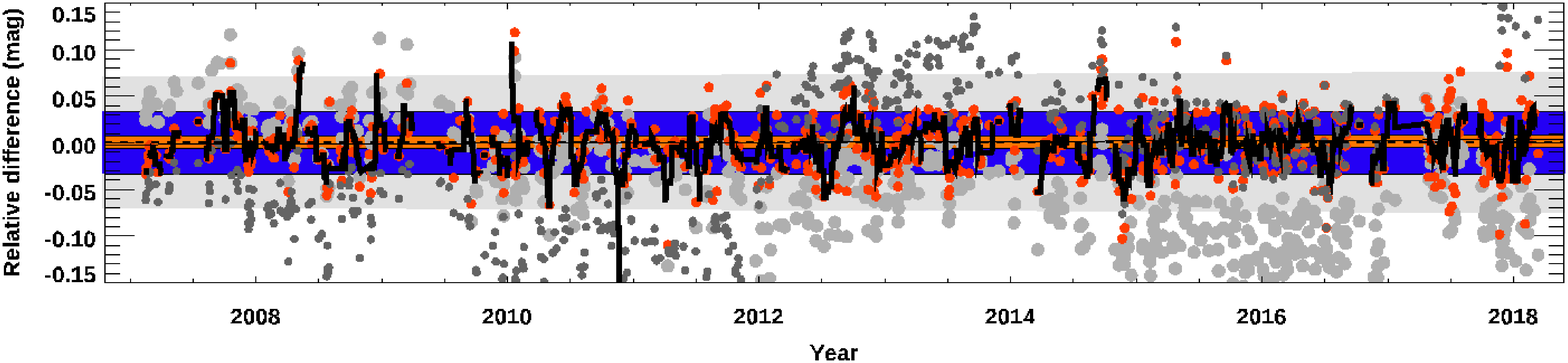}
\caption{GMOS-N ${\rm r}^\prime$ pipeline zeropoint, its median (dashed) and simulations (GMOS-S arbitrarily offset, in grey). Vertical lines indicate detector changes and mirror re-coatings. Below are ${\rm r}^\prime$ pipeline residuals (red; monthly averages: black curves); simulations with two mirror cleanings per year (light-filled circles) or just one (dark); maximal range of model (grey shading), minimum (blue) and mean slope (orange).}
\label{figure_zeropoint_models_with_time}
\end{figure*}

The observed magnitude $m$ of a flat-spectrum object with flux $f_\nu$ (per unit frequency $\nu$) is relative to zeropoint $$m-m_{\rm zeropoint}=2.5\log{[f_\nu F(t)/{F_0}]}-A, \eqno(5)$$ where $F=F_{\rm mirrors}\times F_{\rm optics}\times F_{\rm filter}\times F_{\rm detector}$ are calibration factors of all components in the optical train, and $F_0$ is a constant. Under photometric skies, each $m$ differs from its catalog (outside of atmosphere) value as $$m_{\rm difference}=m-m_{\rm catalog}=A, \eqno(6)$$ for sources of fixed brightness, shown in Figure~\ref{figure_model_atmospheric_extinction_distribution} in the median case as a thick white curve; notice that this is the {\it same} distribution (mode and median) as that induced by air plus cloud. Thus, reformatting using equation 4, it is $${\hat A}={\tilde A}+ (m_{\rm difference}-m_0)-{\tilde m}_{\rm zeropoint}-A_{\rm offset} \eqno(7)$$ against the medians of extinction and zeropoint; $m_0$ is set by the catalog photometric system and filter, 
although need not be known if only relative differences are sought.

Herein lies the utility of a whole-sky catalog such as {\it Gaia} \citep[DR2 is used here]{Gaia2018} on primarily Galactic stars, which at depth of 21 mag in G means that essentially any GMOS imaging observation includes several (later shown to be $\sim 10$ or more) avoiding the need to visit specific standard starfields. The SDSS \citep{York2000} is not all-sky, so cannot fill that role on its own, but overlapping coverage down to 22 mag in ${\rm r}^\prime$ can cross-calibrate North and South with {\it Gaia}; Figure~\ref{figure_model_atmospheric_extinction_distribution} indicates the difference between {\it Gaia}-G and SDSS ${\rm r}^\prime$ catalog magnitudes (white region, to be found directly from the catalogs themselves in Section~\ref{photometry}). And although this ``residual" is somewhat asymmetric, its peak is narrow; notice that near the apex it is about as sharp as the photometric-error limit. So, the assumptions of spectral flatness and of a fixed object-brightness distribution will become safer with more samples, incurring incrementally less than a percent-level error in the peak by subtracting the peak-to-median difference $m_{\rm {\it Gaia}, 0} - m_{\rm SDSS, 0}\approx 0.0040~{\rm mag}$. Equation 7 can therefore allow an estimate of the current atmospheric attenuation relative to its median when instrumental fluctuations are also sufficiently well sampled against (uncorrelated) seasonal timescales.

\begin{figure}
\plotoneextratwiddle{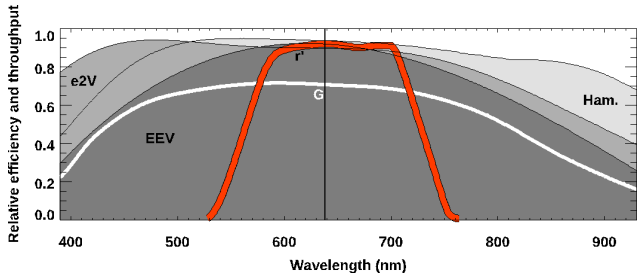}
\caption{{\it Gaia} G throughput (white curve) and the Gemini GMOS ${\rm r}^\prime$ Sloan filter (combined for North and South) overplotted on detector efficiencies (shaded); vertical line: maximal overlap.}
\label{figure_calibration_efficiencies_and_responses}
\end{figure}

A numerical simulation was developed to investigate the required sampling. The dates and estimated throughput of each Gemini mirror recoating and GMOS detector change makes that straightforward. Those changes are indicated by labels in Figure~\ref{figure_zeropoint_models_with_time}. All the optics are identical, and the combined (average of N and S) ${\rm r}^\prime$ filter is shown in Figure~\ref{figure_calibration_efficiencies_and_responses} as a function of wavelength; these are taken from the instrument description\footnote{http://www.gemini.edu/instrumentation/current-instruments} either obtained during commissioning, or discussed in \cite{Jorgensen2009}. The {\it Gaia} bandpass is that measured pre-launch \citep{Weiler2018}. The GMOS detector throughputs are shown for comparison; dark grey indicates the original EEV ones at commissioning (with which roughly half of the data will be obtained); progressively lighter shading indicates the EEV-upgrade performed on GMOS-S (hereafter S-EEV), the E2V chips installed in GMOS-N (N-e2V) and the latest replacement with Hamamatsu devices (N- and S-Ham.). These differences are minimized within ${\rm r}^\prime$ at $638~{\rm nm}$, where the response of {\it Gaia} is also flattest. Folded into that are the dates of mirror recoatings, either primary (M1), secondary (M2), or the tertiary or science fold (SF) which feeds each GMOS. The presciptions of \cite{Schneider2016} and \cite{Schneider2018} were used; these assume that (without cleaning) each of the three coatings decays at a constant rate of $2\%$ per year starting from an initial reflectivity of $95\%$. The effect of cleaning optics was incorporated by lessenning the rate (by half per cleaning) and re-calculating based on whether that was done either once or twice per year. No degradation of the internal optics of the instruments are included although those would also be slow, likely to be comparable to other optics in the train. Likewise, other potential minor instrument misalignments or drifts in central filter-wavelengths are not included; also expected to be small relative to the changes in the telescope optics.

Gemini provides ``pipeline" calibration GMOS-N zeropoint\footnote{http://www.gemini.edu/sciops/instruments/ipm/data-products/gmos-n-and-s/photometric-zeropoints.html} estimates, with which the simulation can be compared. They are obtained from photometry of images in standard fields when skies are clear, after data-pipeline processing of bias, gain, flatfield and background subtraction. The zeropoint is reported from a fit obtained over the interval between re-coatings or detector changes, so it cannot distinguish that from atmospheric change over the same time period. However, the overall ${\rm r}^\prime$ medians of $28.329~{\rm mag}$ (North) and $28.342~{\rm mag}$ (South) will serve as the benchmark. From that, these have typically under $5\%$ residual scatter over timescales of about a year, although some periods (such as in 2009) samples were obtained fewer than once per month (with a total of 3,751 values over 11.1 years) shown in Figure~\ref{figure_zeropoint_models_with_time}: a subscripted label (either 1, 2 or 3) is the recoat after the (initial) one present at first light. One cleaning of an internal optic for GMOS-S was undertaken in 2017, and further M1, M2 and SF re-coatings occured after 2018, some of which were delayed to 2021, and so those are outside the time that will be analyzed \citep{Adamson2022}. Simulated GMOS-N zeropoints are shown in Figure~\ref{figure_zeropoint_models_with_time}. That the simulation, which incorporates optics and detector but no atmosphere, is a good match to the pipeline value suggests that to reach $2\%$ precision relative to catalogs per year - and so discriminate atmospheric from known instrumental changes - sampling and/or accuracy should be improved by a factor of about $(5\%/2\%)^2\approx 6$. For 10 or more 20\%-error samples instead spread uniformly over 20 years, or about $6\times 3,751=22,506$ image frames, the peak in object-magnitude distribution (if narrow and fixed) might be refined to $0.20/\sqrt{225,060}\approx 0.0004~{\rm mag}$. Put another way, the per-frame uncertainty may be closer to $10\%$, but sufficient sampling can quickly beat that error down below 2\%, to within an accuracy that could detect a loss of throughput not accounted for in the optical model. Thus, by obtaining about 1,000 clear-sky frames per year, there can be roughly millimag-level sensitivity to attenuation changes on a per-decade timescale.

\subsection{Prescribing the Method, and Expected Result}\label{method}

The goal is to obtain sufficient samples uncorrelated with either instrument changes or fluctuations due to varying sky conditions, particularly cloudiness. All will be anchored to the combined GMOS benchmark median zeropoint of $28.3355~{\rm mag}$, for which the associated overall median in object versus SDSS ${\rm r}^\prime$ catalog magnitude difference, or $\tilde A$, is called the ``extinction."  This is the conventional meaning (as in equation 5) that includes cloud, directly comparable to Observatory results; the output (for those frames which include SDSS sources) is the current GMOS zeropoint, by definition. Every observation has {\it Gaia}-G samples, returning a ``proxy" zeropoint: ${\tilde m}_{\rm zeropoint} - m_{\rm difference, {\it Gaia}} - (m_{\rm {\it Gaia}, 0} - m_{\rm SDSS, 0})$, and the current peak attenuation estimates for each frame: $${\hat A}={\tilde A}+m_{\rm difference, {\it Gaia}}-{\tilde m}_{\rm difference, {\it Gaia}}-A_{\rm offset}. \eqno(8)$$ As the underlying catalog distributions do not change, and possible contamination by cloud is included in a constant offset, the accumulated sample of those estimates binned by year should average out seasonal variation and allow sensing a change in ${\hat A}$.  Hereafter this is referred to as ``attenuation,"  distinguishing it from extinction by excluding cloud, and can be expected to shift by at least $0.5~{\rm mmag}/{\rm decade}$ to at most $3.1~{\rm mmag}/{\rm decade}$, skewed by rising global air temperature.

\section{Data, Reductions and Analysis}\label{analysis}

\begin{figure}
\plotone{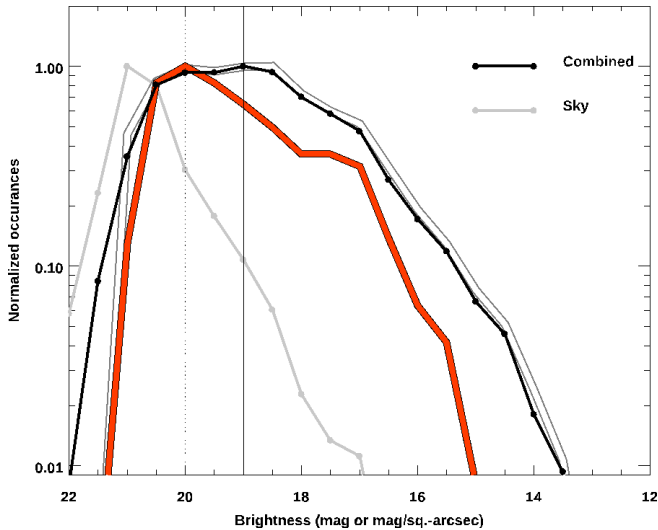}
\caption{Histograms of GMOS ${\rm r}^\prime$ brightness for every object identified from the {\it Gaia} catalog (white curve), SDSS (red) and the total of those samples when combined (black); grey curve indicates the distribution of sky-background brightness values; a 5-sigma point-source sensitivity is indicated by the vertical dotted line; peak of the sample distribution is at $G=19.0$ mag.}
\label{figure_magnitude_histogram}
\end{figure}

\begin{figure*}
\plottwocolumnmedium{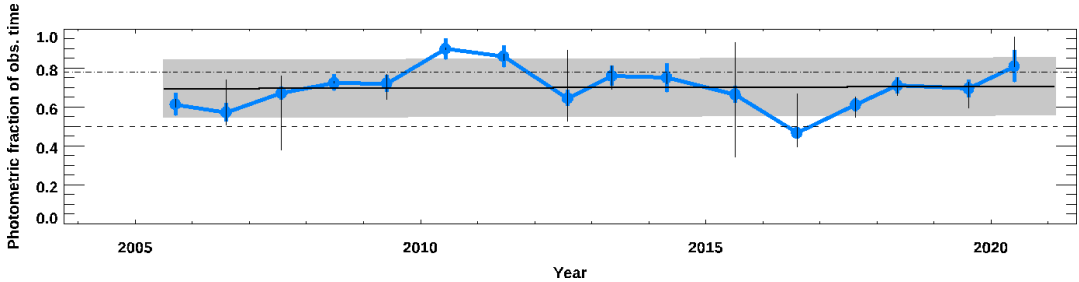}
\caption{GMOS photometric fraction of observations, averaged in time over each calendar year for both; error bars are standard deviations per bin, whiskers are limits of North and South; values given in Table~\ref{samples}. The solid line is least-squares fit, within a $10\%$ range shaded grey.}
\label{figure_photometric_fraction_with_time}
\end{figure*}

\begin{deluxetable*}{lrrrccccccc}
\tablecaption{Gemini Observations and Fractions of Time\label{samples}}
\tablewidth{0pt}
\tabletypesize{\tiny}
\tablehead{\colhead{} &\multicolumn{3}{c}{} & &\multicolumn{6}{c}{Observed Fractions}\\
\cline{6-11}\\
\colhead{} &\multicolumn{3}{c}{} & &\multicolumn{2}{c}{} & &\multicolumn{3}{c}{Sky Conditions}\\
\cline{9-11}\\
\vspace{-4.25 mm}\\
\colhead{} &\multicolumn{3}{c}{GMOS ${\rm r}^\prime$ {\it Gaia}-G Samples} & &\multicolumn{2}{c}{Combined SDSS} & &\multicolumn{2}{c}{Photometric} &\colhead{}\\
\cline{2-4}\cline{6-7}\cline{9-10}\\
\vspace{-5.25 mm}\\
\colhead{Year} &\colhead{North} &\colhead{South} &\colhead{Both} & &\colhead{North} &\colhead{South} & &\colhead{North} &\colhead{South} &\colhead{Best}}
\startdata
2005.7 & 6,672 & 4,326 &10,998 & &0.29 &0.02 & &0.60 &0.64 &0.54\\
2006.6 & 4,098 &10,836 &14,934 & &0.27 &0.05 & &0.74 &0.51 &0.45\\
2007.6 & 9,047 &39,422 &39,422 & &0.29 &0.15 & &0.38 &0.76 &0.58\\ 
2008.5 & 2,834 &26,865 &29,699 & &0.56 &0.09 & &0.73 &0.72 &0.35\\
2009.4 & 9,711 &17,652 &27,363 & &0.34 &0.11 & &0.64 &0.77 &0.67\\
2010.4 & 4,266 &22,613 &26,879 & &0.26 &0.11 & &0.90 &0.90 &0.57\\
2011.5 & 6,807 &16,754 &23,561 & &0.43 &0.17 & &0.84 &0.87 &0.72\\
2012.6 & 7,450 &15,775 &23,225 & &0.26 &0.04 & &0.89 &0.53 &0.50\\
2013.4 &12,720 & 9,634 &22,354 & &0.50 &0.03 & &0.81 &0.69 &0.66\\
2014.3 & 5,996 & 4,307 &10,303 & &0.51 &0.03 & &0.77 &0.72 &0.66\\
2015.5 &12,430 &10,297 &22,727 & &0.38 &0.01 & &0.93 &0.35 &0.60\\
2016.6 &21,220 & 7,928 &29,148 & &0.23 &0.06 & &0.39 &0.67 &0.39\\
2017.6 &12,677 &19,887 &32,564 & &0.64 &0.08 & &0.55 &0.65 &0.39\\
2018.4 &11,702 &16,097 &27,799 & &0.18 &0.02 & &0.65 &0.75 &0.45\\
2019.6 & 5,490 &18,109 &23,599 & &0.21 &0.04 & &0.59 &0.73 &0.39\\
2020.4 & 9,406 &   387 & 9,793 & &0.22 &0.25 & &0.80 &0.96 &0.68\\
\cline{1-11}\\
\vspace{-5.25 mm}\\
Total: &142,802 &231,844 &374,643 & &0.34 &0.08 & &0.78 &0.78 &0.67
\enddata
\end{deluxetable*}

All available GMOS ${\rm r}^\prime$ pipeline-processed images were downloaded from the Gemini Archive, in addition to every ${\rm r}^\prime$-filter exposure designated in its header as taken for ``science."  Although ${\rm r}^\prime$ data was used exclusively in further analysis, it constitutes 48\% of all imaging for science obtained with GMOS in the four main SDSS filters (${\rm g}^\prime$, ${\rm r}^\prime$, ${\rm i}^\prime$ and ${\rm z}^\prime$). Header information prior to 2005 was sometimes incomplete, and those frames were not used. Science observations are on intended targets, excluding commissioning, acquisition, alignment, or any other frame obtained for calibration, e.g. twilight flats. Those others were usually found to be of too short exposure for sufficient signal in the analysis anyway. Good exposures were typically less than $90~{\rm s}$; the mean was $123~{\rm s}$; a few of $600~{\rm s}$ or longer were obtained, but none was retained due to saturation. Only publicly-accessible data were used, and because the standard proprietary periods are between 6 to 12 months, a uniform dataset (of every ${\rm r}^\prime$ image taken) concluded in 2021.

Reduction of all non-pipeline data employed the DRAGONS software\footnote{DRAGONS v3.0.1 available from www.gemini.edu.} for basic bias subtraction, gain, and flatfielding correction. No significant systematic differences with the pipeline-processed ${\rm r}^\prime$ frames were found. A custom IDL (Interactive Data Language) software located objects and performed photometry in each frame. The file-header sky-position angle, and centroid of brightest star (often used for guiding, and so vignetted and unusable) plus one or two fainter stars were used to orient the frame and identify all objects in the field with the Vizier database, and the SDSS and {\it Gaia} catalogues.

\subsection{Photometry and Samples}\label{photometry}

Photometry was carried out on every identified object using a $4~{\rm arcsec}$ aperture, surrounded by a $2~{\rm arcsec}$-wide annulus for obtaining the sky background. By inspection, this aperture was found to be sufficiently large for magnitude 21 pointsources, even under poorer seeing conditions; fitting with a Moffat profile eliminated those either too large or of otherwise-extended objects. All results were then corrected to zenith airmass. Figure~\ref{figure_magnitude_histogram} shows histograms of object brightnesses for all identified {\it Gaia} and SDSS objects. (The ``residual" between catalogs was also recorded; see Section~\ref{instrument}.) The combined result is shown as a thick black curve; the grey curve indicates the measured sky brightnesses, and the thin verical line marks the peak of the distribution; an effective point-source limit is two magnitudes fainter.

There were a total of 1,036,770 samples obtained, of which 288,043 were saturated, 329,401 were faulty by other means (cloud of $1.0~{\rm mag}$ or thicker; misidentifications causing misalignment of the frames, sometimes resulting in the wrong amplifier gain being obtained from the header; others were bad because the object was either vignetted by the guider probe arm, fell too near the edge of the detector array or in a chip gap) with 109,115 broader than the aperture (usually star-trails due to non-siderial guiding) and 145,154 were too noisy (with a signal-to-noise ratio under 5). A final 374,769 remained when requiring no fewer than 5 independent samples per frame (just sufficient to determine a suitably robust median). Some frames had over 100 samples, typically $\sim 20$, spread over 17,812 frames from 2004 to 2021, spanning 17 years; tallied in Table~\ref{samples}. Usually those provided about 5,000 samples per year per telescope. Gemini South in 2020 is the only significant outlier, although together with the North (and despite COVID-19 related shutdowns) there were still nearly 10,000 samples obtained in that year. As SDSS is a northern survey it provides fewer catalogued objects visible from Gemini South than North; fractions are those frames combining at least one SDSS object, against which the overall median extinction is set. All frames were further defined by sky conditions, discussed in the following section.

\subsection{Cloudiness, Sky-Brightness, Airmass and Seeing}\label{cloudiness}

Gemini image headers include weatherstation meteorological data and a data-quality catagorization: sky clarity as the best $50\%$, $70\%$ or in any conditions the dome was open (CC50, CC70, or Any); sky brightness as dark, grey or bright (SB50, SB80 or Any); precipitable water vapour (WV50, WV80 or Any); and image quality in similar fractional bins (IQ20, IQ70, IQ85 or Any) which are the standardized ones employed data-quality assessment\footnote{http://www.gemini.edu/observing/telescopes-and-sites/sites}. Conditions considered photometric (CC50) had no detectable clouds, which are recorded by year in Table~\ref{samples}, and plotted in Figure~\ref{figure_photometric_fraction_with_time}: the photometric observered fraction for both telescopes is stable to within about 10\%.

\begin{figure}
\plotonemedium{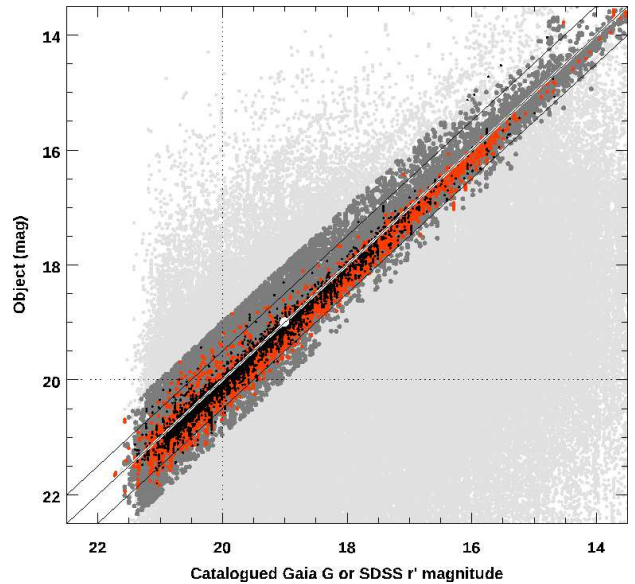}
\caption{Observed object versus catalog magnitudes (dark-grey filled circles: {\it Gaia} G; red: SDSS ${\rm  r}^\prime$, superimposed black dot for South). Excluded (light grey) are those brighter due to faults in photometry, either misidentfication or incorrect gain in header (some faint parallel striping) or fainter, due primarily to cloud. Photometric conditions lie along the 1:1 line; peak of distribution at white dot; point-source photometry limits by dotted lines.}
\label{figure_object_versus_catalog_magnitudes}
\end{figure}

The benign effect of the signal-to-noise cut is illustrated in Figure~\ref{figure_object_versus_catalog_magnitudes}, which is observed object brightness versus the catalog, restricted to CC50. Brighter cases are due primarily to faults in photometry (e.g. gain mismatch) with those fainter reliably associated with visibly cloudy conditions: note tendency for fewer dark-grey dots below the 1:1 line than above - the white line indicates the minimum residual catalog SDSS ${\rm r}^\prime$ versus {\it Gaia} G difference. Whether samples collectively shift (downward) relative to this line is to be looked for.

\begin{figure*}
\plottwocolumnfiddle{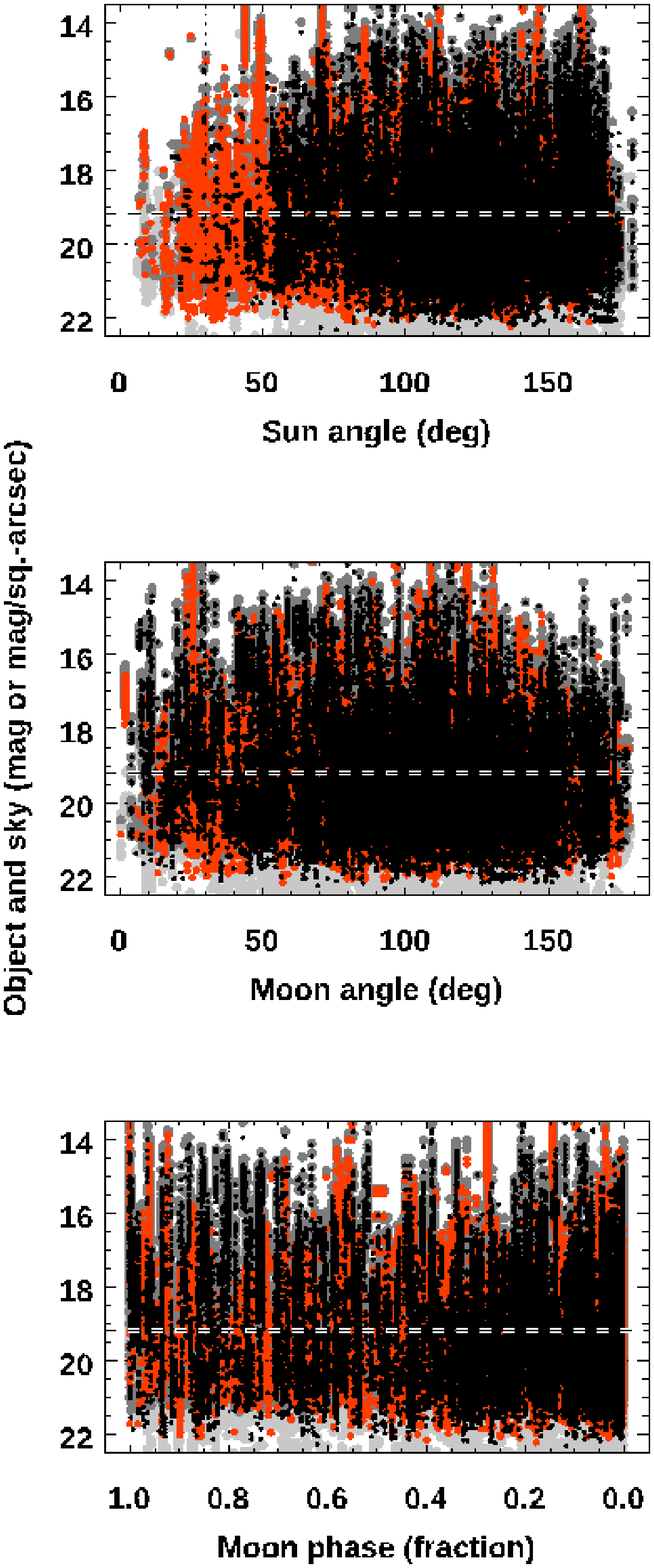}{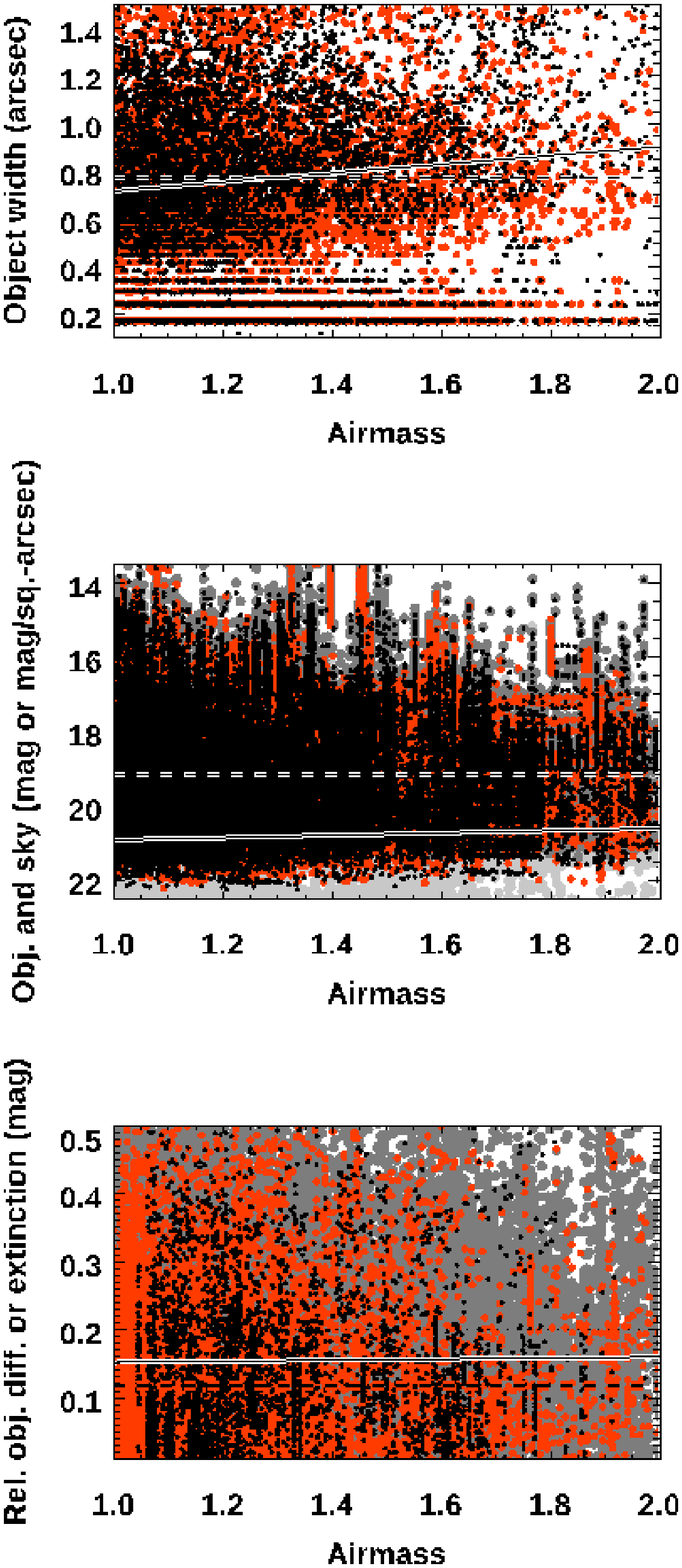}
\caption{Object (red: both North and South; black dot: South) and sky brightness (light grey) with Sun and Moon elevation and Moon phase (left) and airmass (right); object versus {\it Gaia} G and SDSS ${\rm r}^\prime$ difference symbols (lower-right) as in Figure~\ref{figure_object_versus_catalog_magnitudes}; medians are indicated by dashed lines, least-square fits by thin solid lines.}
\label{figure_object_and_sky_with_airmass}
\end{figure*}

No significant bias from sky brightness is considered to affect the results; observations were made at every allowable telescope orientation in all seasons for both. As illustrated in Figure~\ref{figure_object_and_sky_with_airmass}, there seems to be no strong effect with either Sun angle or illumination by the Moon. The dashed lines are median values for object and sky brightness. A linear least-squares fit to each of those relations is indicated by a solid black line.

\begin{figure}
\plotone{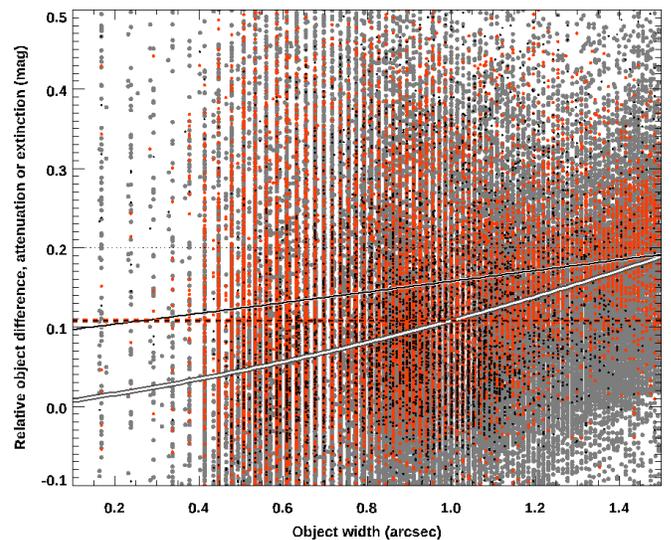}
\caption{Object versus {\it Gaia} G difference (dark-grey filled circles) and extinction in ${\rm r}^\prime$ (red: both North and South; black dot: South) with object width. Median is indicated by red dashed line; least-squares linear fit, thin black line; quadratic: white curve.}
\label{figure_extinction_versus_seeing}
\end{figure}

Unsurprisingly, best photometry was obtained at lowest airmass, where object width was minimized; that is shown in Figure~\ref{figure_extinction_versus_seeing}. A linear relationship would be expected if extinction and atmospheric attenuation were due soley to airmass. Instead, a quadratic fit is also shown (white curve), offset to allow passing through zero at Nyquist sampling. This better fit, decreasing with diminishing object surface area, suggests that photometry under ideal seeing conditions will also draw closest to the limiting precision of the catalog (or equivalently the lowest zeropoint), to be considered when looking for any trend in attenuation. This ``best" fraction of the dataset, conditions considered to be both CC50 and IQ70 or better, were 67\% of the total dataset or $0.67\times 374,643\approx 251,000$ samples. Further restricting to driest conditions (WV50 or WV80) was found only to increase fitting errors by cutting the number samples.

\subsection{Resulting Attenuation Changes with Time}

\begin{figure}
\plotone{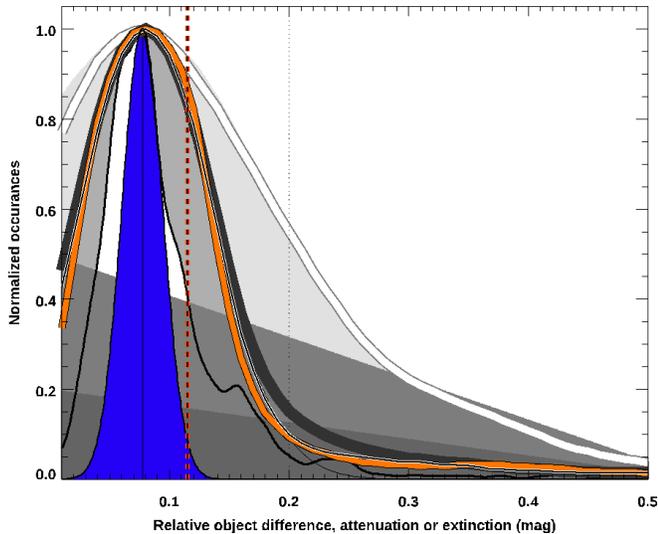}
\caption{Object versus {\it Gaia} G difference (thick white curve), attenuation (thick black) total distributions; the model (shaded regions) are the same as in Figure~\ref{figure_model_atmospheric_extinction_distribution}; vertical red dashed line is overall median SDSS ${\rm r}^\prime$ extinction; orange curve shows distribution for ``worst" period of attenuation (first half of 2009) thin black curve ``initial" dataset under ``best" conditions considered both photometric (CC50) and of image quality IQ70 or better, drawing near the limit of {\it Gaia} G - SDSS ${\rm r}^\prime$ catalog differences (white region).}
\label{figure_difference_and_extinction_distributions}
\end{figure}

Following the prescription of Section~\ref{method} a current estimated atmospheric attenuation was returned in each frame. The total distribution is shown in Figure~\ref{figure_difference_and_extinction_distributions}, with the overall median GMOS SDSS ${\rm r}^\prime$ extinction when photometric ($0.1075~{\rm mag}$) indicated by the vertical dashed red line, or $0.1221\pm 0.0059~{\rm mag}$ (North) and $0.0929\pm 0.0059~{\rm mag}$ (South) for one standard-deviation errors, within uncertainties of the Observatory-reported values of $0.11\pm 0.01~{\rm mag}$ and $0.10\pm 0.01~{\rm mag}$ respectively. And against that, the attenuation does show a change consistent with what was expected from the model, shifting from the minimum possible object-to-catalog difference at the start. An ``initial" dataset ending in 2013 (the first half of data) under ``best" conditions considered to be both CC50 and under IQ70 or better is shown as a thin black curve; the same, but for the ``worst" period of the dataset (the first half of 2009) in orange, with a skew faintward of the distribution peak evident. 

To emphasize: by this method, changes in optical throughput and their differences between North and South are accounted for {\it frame by frame} in accumulating the distributions; a current frame-averaged zeropoint in each (relative to its overall median value) is an output, not an input. Airmass was already corrected for in the photometry, and adding the further complexity of accounting for the barometric pressure-difference between sites in $A_{\rm offset}$ did not improve on the results of a single constant, that is, by narrowing or tightening the distribution of attenuation (not shown in Figure~\ref{figure_difference_and_extinction_distributions}). Likewise, attempting to ``counteract" a drift in pressure by inverting its sense in airmass corrections (but keeping same median) does not remove the skew towards worse attenuation of the peak relative to its median: the distribution shape has changed, which is unaffected by a renormalization. Note that any zeropoint change from frame-to-frame must affect all objects within each frame in the same way, so accuracy should improve with more objects per frame, and to the extent that those sufficiently sample fluctuations month-to-month and year-to-year. Poorer attenuation would be in addition to decay of optical coatings, which is (typically) much steeper, producing ``jagged" changes with time and occasional large differences between sites. But a shifting distribution is consistent with what is seen, rather than a broadening one, as in the latter case it would be expected to also stretch brightward, outside the lower limit of the model envelope (towards the left). Recall that the attenuation estimate here is relative to the object-distribution peak, which is interpreted to have shifted (towards the right, demarcated at the start by the blue faintward ``edge" of the residual difference) pushing the associated {\it Gaia}-derived attenuation (plus constant offset) past its SDSS-set overall median extinction (vertical red dashed line), away from its initial value (and zero). It is safe to assume the sources did not collectively (and inherently) get fainter to cause that; the catalog reference magnitudes were fixed. As the observed conditions were not visibly becoming cloudier on average during this time, the sensed change can be attributed to worsening attenuation by air.

\begin{figure}
\plotone{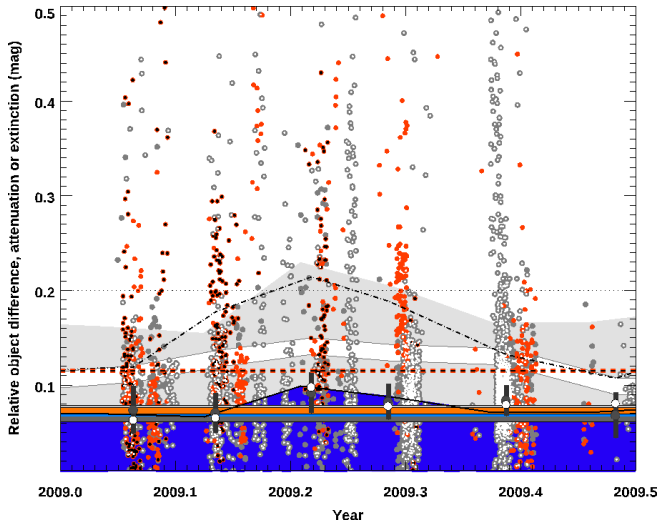}
\caption{Object versus {\it Gaia} G difference (white curve) centred on first half of 2009, and monthly-averaged attenuation (large black filled-circles; best-conditions: white) and SDSS ${\rm r}^\prime$ extinction (thin dot-dashed black curve) together with its overall-dataset median (red dashed line); orange solid line is a fit to all attenuation estimates, overplotted on model of Section~\ref{modeling} and MLO solar radiance measurements (dark-blue region); light-grey filled circles are individual {\it Gaia} samples, white during observer-reported best conditions, randomized (by up to a week) to help show distributions.}
\label{figure_object_difference_attenuation_and_extinction_with_time_zoom}
\end{figure}

The period of worst transparency during this study, in the first half of 2009, is shown in Figure~\ref{figure_object_difference_attenuation_and_extinction_with_time_zoom}, with the monthly-averaged object versus {\it Gaia} G value indicated by the white curve. The dark-blue region is that bounded by the solar radiance measurements from MLO, as was shown in Figure~\ref{figure_attenuation_from_mlo_against_cfht_weather_data}; light grey indicates the same, but amplified by a factor of two. Median SDSS ${\rm r}^\prime$ extinction of the full dataset is again indicated by a (horizontal) dashed red line; the orange line is a least-squares fit to all {\it Gaia}-G estimates (which will be discussed in more detail to follow). The exceptionally low transparency ``spike" of about a month is seen in these attenuation measurements, which is a significant enhancement, but one comparable to that expected over the full duration of the dataset due to longer-term changes. And, in fact, omitting 2009 from a least-squares fit to the atmospheric transparency from MLO data - by excluding those data prior to 2010 - increases that to $4.0~{\rm mmag}/{\rm decade}$. When fitting a dataset overlapping Gemini operations, this gives a slope of $1.7~{\rm mmag}/{\rm decade}$ (see Section~\ref{introduction}), with $A_0$ still within the allowable range of $0.0025~{\rm mag}$ (half the standard deviation) as in Section~\ref{modeling}. All fits are listed in Table~\ref{table_results}; slope uncertainties are given as the standard deviation limits for each.

\begin{deluxetable}{lcccc}
\tablecaption{Attenuation Measurement Fits\label{table_results}}
\tablewidth{0pt}
\tabletypesize{\tiny}
\tablehead{\colhead{} &\colhead{} &\colhead{Sky} &\colhead{$A_0$} &\colhead{$\Delta A$}\\
\colhead{Dataset} &\colhead{Time Period} &\colhead{Conditions} &\colhead{(mag)} &\colhead{(${{\rm mmag}\over{\rm decade}}$)}}
\startdata
\sidehead{MLO Solar Radiance}
Full\footnote{Using all available data.}                                      &1958.0-2021.5 &Clear &0.0694 &$0.4\pm 0.2$\\
Overlap\footnote{Beginning at Gemini first light.}                            &2000.0-2021.5 &Clear &0.0691 &$1.7\pm 0.5$\\
Exclusive\footnote{Excluding data of 2009 episode and prior.}                 &2010.0-2021.5 &Clear &0.0688 &$4.0\pm 1.0$\\
\sidehead{GMOS Photometry}
Initial\footnote{Both telescopes, restricted to first half of available data.}&2005.5-2013.5 &Best  &0.0700 &$1.3\pm 0.9$\\
Both                                                                          &2005.5-2021.0 &Best  &0.0711 &$2.7\pm 1.1$\\
North                                                                         &2005.5-2021.0 &Photometric &0.0698 &$3.9\pm 1.4$\\
South                                                                         &2005.5-2021.0 &Photometric &0.0648 &$4.3\pm 1.4$
\enddata
\end{deluxetable}

\begin{figure*}
\plottwocolumnwide{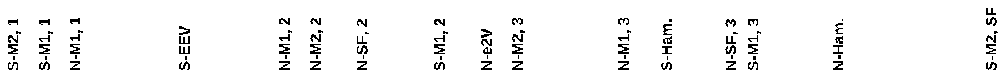}\\
\hspace{1.0 mm}\plottwocolumnwide{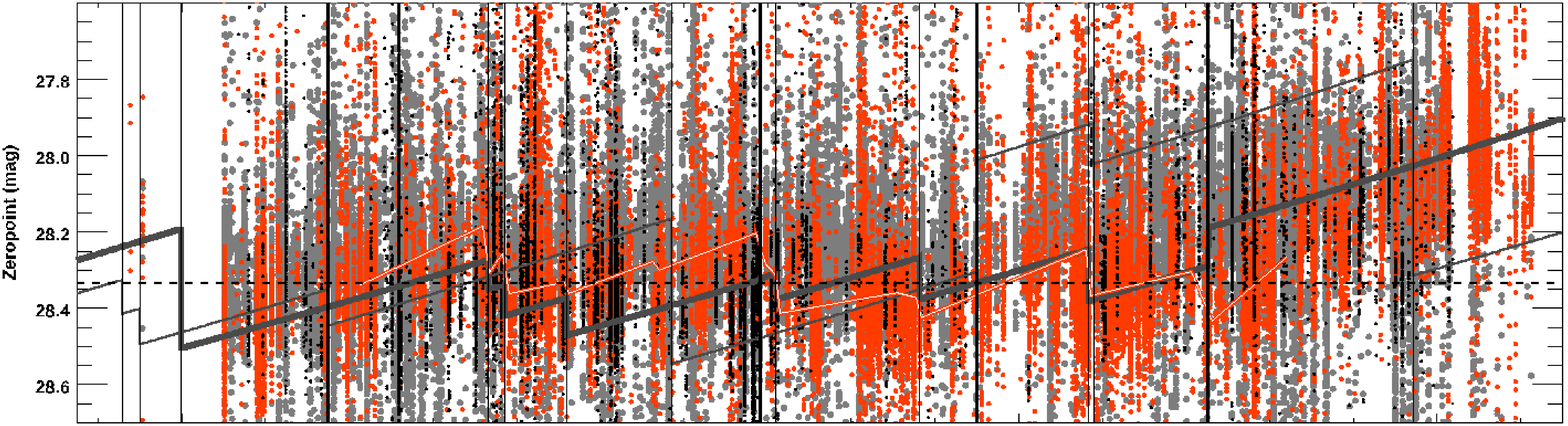}\\
\vspace{1.5 mm}
\hspace{1.0 mm}\plottwocolumnwidetwiddle{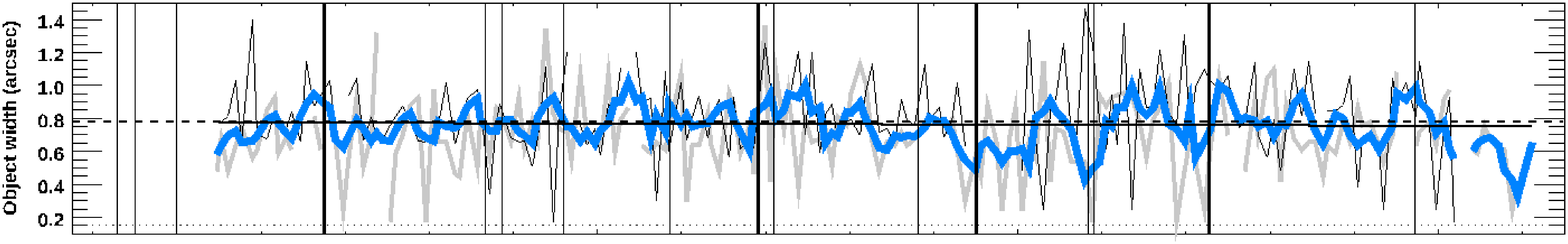}\\
\vspace{1.0 mm}
\hspace{2.0 mm}\plottwocolumnwidetwiddleagain{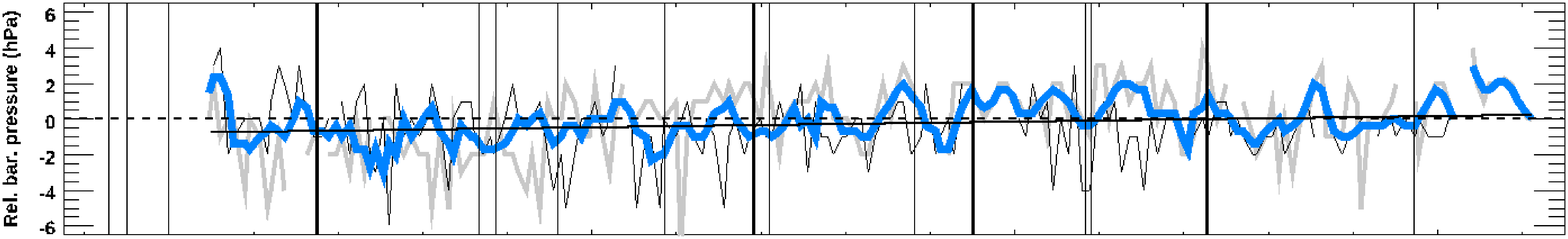}\\
\vspace{1.5 mm}
\hspace{1.0 mm}\plottwocolumnwidetwiddle{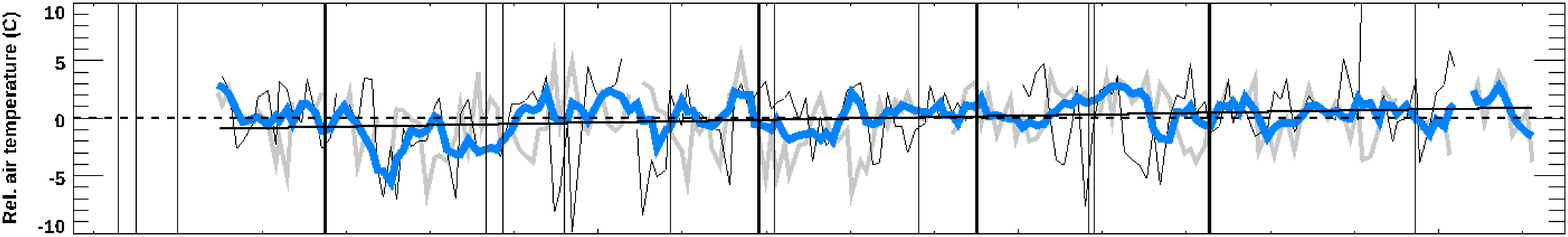}\\
\vspace{1.5 mm}
\hspace{1.5 mm}\plottwocolumnwidetwiddleagain{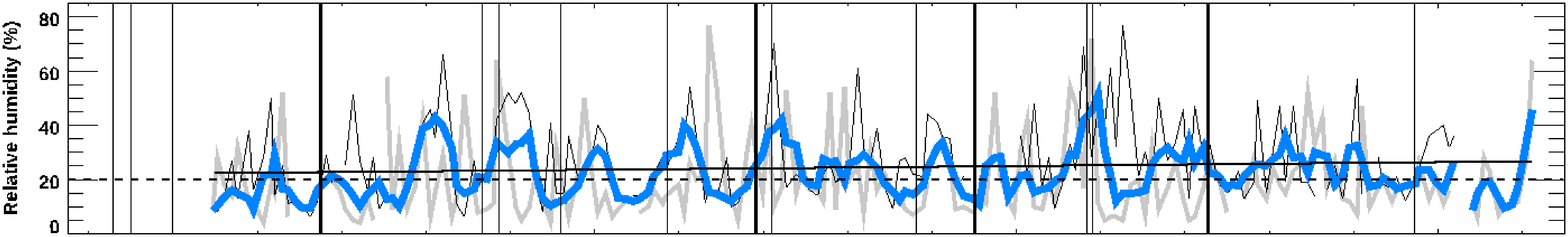}\\
\vspace{1.5 mm}
\hspace{0.5 mm}\plottwocolumnwidetwiddle{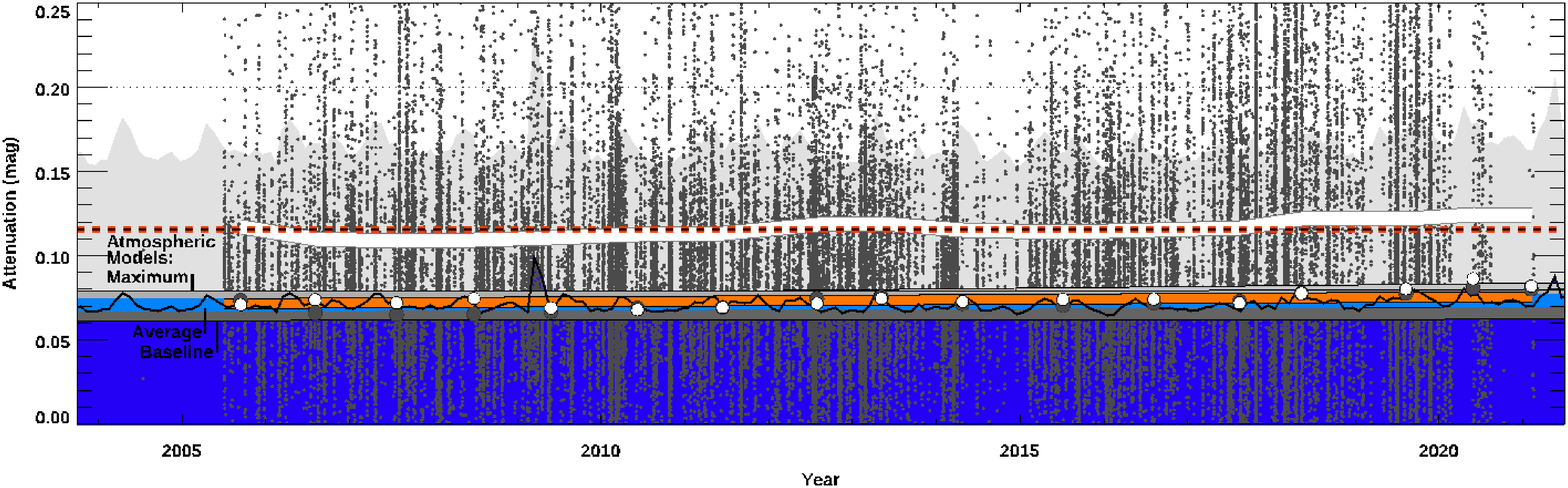}
\caption{Top panel: GMOS ${\rm r}^\prime$ zeropoints (filled red circles; superimposed black dot: South) overplotted on proxies obtained via Gaia objects (filled grey circles) together with simulations from Section~\ref{instrument}; vertical lines indicate major instrument changes; horizontal dashed: overall median. Below: monthly-averaged object widths, barometric pressure, air temperature, and humidity (grey: North, black: South; blue: both, smoothed over 3 months) for all frames shown against medians (dashed lines). Black lines indicate a least-squares linear fit to each. Bottom: attenuation per frame, with yearly averages (large, black filled circles) and restricted to best data during photometric conditions (white) with fit indicated by orange line: an upward trend is within range of the simple model atmosphere (baseline to maximum: shading; average: light blue) and steeper than MLO transparency decrements (thick black curve); a similar tendency is seen in yearly-mean Gaia-G object-difference magnitude (white curve; smoothed over 3 years) against total-sample median SDSS ${\rm r}^\prime$ extinction (dashed red line).}
\label{figure_all_data_with_time}
\end{figure*}

Long-term trends in the Gemini dataset also agree with the MLO measurements, CFHT meteorology, and the simplified-model expectations for air temperature rise. Monthly averages of barometric pressure, surface air temperature, and humidity relative to their medians are shown in Figure~\ref{figure_all_data_with_time}. These combine both telescopes when samples were taken, so reflect the observed average conditions for the dataset including a larger fraction of samples from the South. Least-square fits of each give slopes per decade that are all increasing: pressure by $0.66~{\rm hPa}$, temperature by $1.3~^\circ {\rm C}$, and relative humidity by $3.0\%$, giving an incremental constant closely matching the CFHT-set value in equation 3 over this time ($2.22~{\rm mmag}~^\circ {\rm C}^{-1}$). Also shown is a plot of monthly-averaged object width for which no overall trend (as a slope) is detectable. It is notable, however, that despite a best-recorded value ($0.4~{\rm arcsec}$, effectively the average seeing) for the North occuring in January of 2021, five of the all-time poorest months ($\geq 1~{\rm arcsec}$ on average for both telescopes) occured in the last 5 years. If not due to poorer telescope performance (which is monitored) they could be excursions of bad seeing. Derived zeropoints (relative to the overall median, top panel) are provided together with simulations from Section~\ref{instrument}. The bottom panel is the resultant attenuation relative to overall median extinction (red horizontal dashed line). Although not entirely a monotonic increase, the yearly-averaged object to {\it Gaia}-G catalog magnitude difference (shown offset by $A_{\rm offset}$) tends slowly upward. Poor tracking of zeropoint might be expected to incur discontinuities, but no sharp ``kink" or ``bend" corresponds to any particular major instrumental change. Rather, yearly atmospheric-attenuation averages (plotted at their data-weighted year midpoints) and those restricted to best data during photometric conditions (CC50, IQ70 or better) are within their uncertainties (roughly the size of these symbols) to a linear least-squares fit indicated by an orange line. Photometric-only data (separate for North and South, not shown) give steeper slopes, poorer constrained. Within errors, all match the (decrement in) transparency from MLO aerosol content as well as the solar radiance measurements (thick black curve, with dark-blue shading) together with the average model of Section~\ref{modeling} as indicated by the light-blue region; the same model using Gemini-derived meteorological data would not be visibly different on this plot. Overall, the average model case of higher humidity together with warming air temperature is consistent with growth in attenuation.

\section{Summary and Conclusions}\label{summary}

Archival data from the Gemini GMOS instruments, including almost every Sloan ${\rm r}^\prime$-filter science image obtained with those and comprising over 250,000 samples in near-ideal photometric conditions spanning 17 years, has been used to infer long-term changes in atmospheric attenuation. A prescription for time-dependence of zeropoints is effective in prediction to within a few percent per year, in agreemnent with established pipeline data reductions. This guides the method, and requirements for sufficient samples to overcome those effects and be confident that the atmospheric component alone can be sensed. A brief period of significantly enhanced extinction in 2009 is evident, independent of a longer-term trend of increasing attenuation. Those decrements are of comparable amplitude, however, and sensing them is possible because a single small offset, $A_{\rm offset}$, can return an estimate of ${\hat A}$, accounting for unseen contamination by thin cloud (even under what are considered visually to be the best-possible skies) in observed object median magnitude, and {\it Gaia} provides a catalog of sufficient sources from which to define that in essentially any GMOS science frame obtained. That calibration was pinned to the SDSS photometric system (directly) and in comparison to existing pipeline-reduced standard starfield calbrations, which are regularly obtained by the Observatory. Attenuation worsens over the full dataset by approximately $2~{\rm mmag}/{\rm decade}$ ($\approx 0.20\%/{\rm decade}$) since the start of Gemini operations, in agreement with solar radiometry measurements ($-0.17\%/{\rm decade}$). A plausible model is simply a rise in air temperature, with associated change in pressure and humidity, of about $0.7~^\circ{\rm C}/{\rm decade}$. That is over a factor of three beyond the minimum expected for post-industrical temperature rise of $0.2~^\circ{\rm C}/{\rm decade}$. This is a concerning result, even with the caveat that the latter value is only a reference point: a surface-level sea-level temperature meant to track overall change. It is hoped to spur further scrutiny of ground-based astronomical data, looking for possible effects of a warming global climate. 

This study has demonstrated that tracking the atmospheric attenuation component at least weekly, and to better than about $2\%$ accuracy per year is necessary to avoid neglecting growing attenuation from the photometric error budget - whether that were from volcanic activity or a trend - which was shown to be just-achieveable with these archival data. It suggests that for future surveys, more frequent and regular standard starfield observations are required. And even though the increase seen here in atmospheric attenuation growth per year may be small, after 23 years it will have accumulated to a nearly 1\% net loss in the sensitivity of Gemini North and South in the optical (combined, the photon throughput would effectively be down $2\times 23\times 0.02\%\approx 0.92\%$). Although this is roughly equivalent to the benefit derived from cleaning a mirror (just once) it cannot be recovered by any amelioration of the optics, and so is inherent in their zeropoints. In concrete terms, as the effect is proportional to primary collecting area, every year each of the two Gemini telescopes loses in essence an equivalent $\sqrt{0.02\%}\times 8.2~{\rm m}$, or a $12~{\rm cm}$ aperture. As this is seemingly a worldwide effect, all OIR telescopes are experiencing the same decremental loss, so compounded by 13 for the current 8-m class observatories; relevant to Rubin Observatory and other, larger upcoming facilities.  

\acknowledgements

Helpful advice from Andy Adamson, kindly sharing his experience and insight on Gemini operations, is gratefully acknowledged; I thank Julia Scharwaechter for thoughtful comments on an early draft, and an anonymous referee for many suggestions which much improved the quality of the final manuscript. This work was based on observations obtained at the international Gemini Observatory, a program of NSF’s NOIRLab, which is managed by the Association of Universities for Research in Astronomy (AURA) under a cooperative agreement with the National Science Foundation on behalf of the Gemini Observatory partnership: the National Science Foundation (United States), National Research Council (Canada), Agencia Nacional de Investigaci\'{o}n y Desarrollo (Chile), Ministerio de Ciencia, Tecnolog\'{i}a e Innovaci\'{o}n (Argentina), Minist\'{e}rio da Ci\^{e}ncia, Tecnologia, Inova\c{c}\~{o}es e Comunica\c{c}\~{o}es (Brazil), and Korea Astronomy and Space Science Institute (Republic of Korea). It was enabled by observations made from the Gemini North telescope, located within the Maunakea Science Reserve and adjacent to the summit of Maunakea. We are grateful for the privilege of observing the Universe from a place that is unique in both its astronomical quality and its cultural significance. These data were acquired through the Gemini Observatory Archive at NSF’s NOIRLab and processed using the DRAGONS (Data Reduction for Astronomy from Gemini Observatory North and South).


\begin{thebibliography}{}

\bibitem[Adamson(2022)]{Adamson2022} Adamson, A. 2022, private communication

\bibitem[Backhouse(1893)]{Backhouse1893} Backhouse, T.W. 1893, Nature, 48, 509

\bibitem[B{\"o}hm et al.(2020)]{Bohm2020} B{\"o}hm, C., Reyers, M., Schween, J.H. \& Crewell, S. 2020, Global and Planetary Change, 190, 103192

\bibitem[Cantalloube et al.(2020)]{Cantalloube2020} Cantalloube, F., Milli, J., B{\o}hm, C. et al. 2020, Nat Astron 4, 826

\bibitem[Coughlin et al.(2018)]{Coughlin2018} Coughlin, M.W., Deustua, S., Guyonnet, A., Mondrik, N., Rice, J.P., Stubbs, C.W. \& Woodward, J.T. 2018, \procspie, 1070420

\bibitem[Dutton et al.(1985)]{Dutton1985} Dutton, E.G., DeLuisi, J.J. \& Austring A.P. 1985, Journal of Atmospheric Chemistry, 3, 53

\bibitem[Dutton \& Bodhaine(2001)]{Dutton2001} Dutton, E.G. \& Bodhaine, B.A. 2001, Journal of Climate, 1409, 3255

\bibitem[Dutton et al.(2011)]{Dutton2011} Dutton, E.G., Daniel, J.S., Neely, R.R. III, Vernier, J.-P., Dutton, E.G. \& Thomason, L.W. 2011, 333, 866

\bibitem[Flagey et al.(2021)]{Flagey2021} Flagey, N., Thronas, K., Petric, A.O., Withington, K. \& Seidel, M.J. 2021, J. Astron. Telesc. Instrum. Syst., 7(1), 017001-1

\bibitem[Gaia Collaboration(2018)]{Gaia2018} {\it Gaia} Collaboration, 2018, Astron. \& Astrophys., 616, A1

\bibitem[Haslebacher(2022)]{Haslebacher2022} Haslebacher C., Demory, M.-E. , Demory B.-O., Sarazin, M. \& Vidale, P.L. Astron. \& Astrophys., 665, A149

\bibitem[Hook et al.(2004)]{Hook2004} Hook, I.M., Jorgensen, I., Allington-Smith, J.R., Davies, R.L., Metcalfe, N., Murowinski, R.G. \& Crampton, D. 2004, \pasp, 116, 425 

\bibitem[Jorgensen(2009)]{Jorgensen2009} Jorgensen, I. 2009, \pasa, 26, 17

\bibitem[Miles(1983)]{Miles1983} Miles, R., 1983, Journal of the British Astronomical Association, 93, 5, 233

\bibitem[Murowinski et al.(1998)]{Murowinski1998} Murowinski, R.G., Bond, T., Crampton, D., Davidge, T.J. et al. 1998, Proc. SPIE Vol. 3355, p. 188-195, Optical Astronomical Instrumentation, Sandro D'Odorico; Ed.


\bibitem[NOAA(2021)]{NOAA2021} NOAA National Centers for Environmental Information, State of the Climate: Monthly Global Climate Report for 2020, published online January 2021

\bibitem[Oke \& Gunn(1983)]{Oke&Gunn1983} Oke, J.B. \& Gunn, J.E. 1983, \apj, 266, 713

\bibitem[Sarazin et al.(2008)]{Sarazin2008} Sarazin, M., Melnick, J., Navarrete, J. \& Lombardi, G. 2008, Messenger, 132, 11

\bibitem[Schneider et al.(2016)]{Schneider2016} Schneider, T., Vucina, T., Ah Hee, C., Araya, C. \& Moreno, Proc. SPIE 9906, Ground-based and Airborne Telescopes VI, 990632

\bibitem[Schneider \& Stupik(2018)]{Schneider2018} Schneider, T. \& Stupik, P. 2018, Proc. SPIE 10700, Ground-based and Airborne Telescopes VII, 1070048

\bibitem[Steinbring et al.(2009)]{Steinbring2009} Steinbring, E., Cuillandre, J.-C. \& Magnier, E. 2009, \pasp, 121, 295

\bibitem[Steinbring et al.(2012)]{Steinbring2012} Steinbring, E., Ward, W. \& Drummond, J.R. 2012, \pasp, 124, 185

\bibitem[van Kooten \& Izett(2022)]{vanKooten2022} van Kooten, M.A.M. \& Izett, J.G. 2022, \pasp, {\it accepted}

\bibitem[Weiler(2018)]{Weiler2018} Weiler, M. 2018, Astron. \& Astrophys., 617, A138

\bibitem[York(2000)]{York2000} York, D.G., Adelman, J., Anderson, J.E. Jr., Anderson, S.F. et al. 2000, \aj, 120, 1579

\end{thebibliography}
\end{document}